# Resolving stress state at crack tip to elucidate nature of elastomeric fracture

Zehao Fan and Shi-Qing Wang[*]
School of Polymer Science and Polymer Engineering, University of Akron
Akron, Ohio 44325

**Abstract**

Based on spatial-temporal resolved measurements of the stress field at crack tip based on polarized optical microscopy (*str*-POM), the stress analysis approach to elastomeric fracture uncovers new insights. We show new phenomenology in contrast to the standard description of linear elastic fracture mechanics (LEFM). First, *str*-POM measurements show emergence of a stress saturation zone whose dimension $r_{ss}$ is independent of the stress intensity factor K. This elastic zone is plastic zone whose size would scale quadratically with K. The absence of stress divergence allows us to measure tip stress $\sigma_{tip}$ at the onset of fracture, identified as inherent material strength, i.e., $\sigma_{tip(F)} = \sigma_{F(inh)}$. We are able to explain why LEFM applies well to elastomers, i.e., why toughness (either given as critical energy release rate $G_c$ or critical stress intensity factor $K_c$) is a material constant, and we have identified parameters that determine the magnitude of toughness. Second, the popular Rivlin-Thomas energy balance description of elastomeric fracture in pure shear has acquired a fresh and different interpretation based on *str*-POM observations, which show that the stress buildup at cut tip explicitly scales with specimen height $h_0$, leading to to $G_c = w_c h_0$ being constant. Third, the *str*-POM observations reveal how elastomeric fracture occurs at a common $K_c$ independent of specimen thickness. At a given load there is weaker stress buildup for a thicker specimen due to greater stress saturation at cut tip, and fracture is observed to occur at lower tip stress for a thicker specimen.

---

[*] Corresponding author at swang@uakron.edu

# I. Introduction

Fracture behavior of solids is a most important topic in materials science and engineering. Soft materials also undergo fracture.[1, 2] After a hundred years[3] of extensive research[4] fracture mechanics has become a more mature subject, guiding research on fracture of modern materials including polymers. Effects of large cracks on the strength of brittle polymers including plastics and elastomers can indeed be characterized by toughness $G_c$, also known as the critical energy release rate.[4, 5] For both glassy polymers[6, 7] and crosslinked rubbers,[8] since $G_c$ has been found to be hundreds or thousands times greater than the surface fracture energy $\Gamma$, within fracture mechanics we can however neither *a priori* prescribe the magnitude of $G_c$ nor rationalize why $G_c$ is a material constant. While any microscopic modeling of $G_c$ can be regarded as a theoretical inquiry to provide a theoretical explanation of why linear elastic fracture mechanics (LEFM) applies, we often fail to find the calculated $G_c$ in agreement with experiment and cannot explain why $G_c$ actually varies by a factor of three[6] for poly(methyl methacrylate). Related to these unanswered questions is whether polymers are flaw intolerant, i.e., whether brittle fracture of polymers without intentional through-cut is due to presence of intrinsic flaws or defects, as suggested.[5]

Since the beginning of LEFM,[9-11] there has been the recognition that "fracture is governed by the local stress and deformation conditions around the crack tip".[12] However, this second pillar of stress intensification analysis in LEFM met the difficulty of predicted stress singularity[10, 13] and quickly folded back to the first pillar, i.e., Griffith[3]-Irwin[14] energy balance argument, thanks to Irwin's demonstration[11, 15] to show equivalence between the two pillars within LEFM: $K_c = (G_c E)^{1/2}$, where $K_c$ is Irwin's critical stress intensity factor and E the Young's modulus. Thus, fracture behavior has ever since been conveniently characterized in terms of global conditions, i.e., through the measurement of $G_c$ based on its operational definition ($G_c \sim \sigma_c^2 a/E$) for a given cut size *a* at critical far-field load $\sigma_c$ for fracture. Until recently[16] fracture of rubber has been exclusively discussed on the basis of Griffith toughness $G_c$.

Separately, according to textbooks[4, 5] $K_c$ changes with specimen thickness for certain materials (e.g., metals). As the state of stress changes from plane stress (thin) to plane strain (thick), $K_c$ tends to decrease.[17] This thickness dependence of $K_c$ acquired a particularly convenient interpretation in the energy approach: there is less plastic dissipation under plane strain, with lower $G_c$ implying lower $K_c$. However, we note that the literature on fracture of rubbers shows[18] independence of fracture toughness on specimen thickness. Moreover, toughness of rubber has been observed to increase with crack propagation speed, examined either in terms of tensile extension of precut specimens[18, 19] or tearing.[20, 21] Since there is little plastic dissipation in rubbers, this correlation between toughness and speed has been suggested to arise from increased molecular friction,[21-23] similar to[24] the rise of loss modulus $\mu''(\omega)$ with frequency $\omega$ in linear viscoelastic characterization of rubber-glass transition.

Based on fracture behavior of two elastomers, in this work we explore and establish that elastomeric fracture can be more conveniently characterized in terms of the crack tip stress exceeding inherent strength, rather than an energy-based argument. Specifically, using recently adopted spatial-temporal resolved polarized-optical microscopy (*str*-POM)[16] we address the following questions to demonstrate the merit of quantifying local processes at crack tip: (a) Is onset of fracture determined by critical stress state at crack tip? (b) Why is $K_c$ a material specific constant and what determines its magnitude? (c) Does Rivlin-Thomas pure-shear protocol for elastomers[8] to measure $G_c$ have a different origin? (d) How does $K_c$ depend on specimen thickness as plane stress changes to plane strain in the case of elastomers? To this end, we need spatial resolution better then 20 µm, which can be accomplished based on 4K video camera coupled to microscopic lens of sufficient high magnification. Because the stress buildup at the tip is found to saturate to a plateau, we are able to show that (1) elastomeric fracture amounts to the stress at crack tip $\sigma_{tip}$ reaching a threshold value $\sigma_{F(inh)}$ that may be identified as inherent strength, (2) $\sigma_{tip}$ grows linearly with the nominal load $\sigma$, or the stress intensity factor operationally defined as[4] $K = 1.12\sigma\sqrt{\pi a}$ (for single-edge-notch), i.e., we find $\sigma_{tip} = K/\sqrt{P}$ to reveal a characteristic length scale P, which originates from stress saturation at crack tip, (3) at fracture $K_c = \sigma_{F(inh)}\sqrt{P}$, (4) in pure shear $\sigma_{tip}$ also scales linearly with nominal load as $\sigma_{tip} = K_{ps}/\sqrt{P}$, with $K_{ps} = \sigma\sqrt{h_0}$, where $h_0$ is the sample height $h_0$, (5) at fracture, the critical stress intensity factor $K_c = \sigma_{F(inh)}\sqrt{P}$ is independent of the sample thickness D (ranging from 0.6 to 5 mm), with $\sigma_{F(inh)}$ decreasing with D and P increasing with D.

Given the *str*-POM observations, we suggest that elastomeric fracture may be understood to arise from the inherent strength $\sigma_{F(inh)}$ being exceeded at crack tip, and the accompanying energy release in terms of $G_c$ is merely a measure of what is involved in fracture. In other words, G → $G_c$ is not the criterion for fracture, but the tip stress $\sigma_{tip}$ → $\sigma_{F(inh)}$ is, which is realized by raising the load level to $\sigma$ → $\sigma_c \sim \sigma_{F(inh)}\sqrt{P/a}$ in the case of single-edge notch (SEN) i.e., the applied load $\sigma$ reaching a certain fraction of $\sigma_{F(inh)}$ that is determined by the ratio of the two lengths, i.e., by P/*a*.

## II. Results

*II.A Fracture of rubber sheets at different cut sizes (SEN)*

Rivlin and Thomas has established[8] that for stretchable elastic bodies such as rubber fracture behavior can be characterized by measuring its toughness $G_{c(RT)}$ in terms of the work density of fracture $w_c$ and cut size *a*:[8, 25, 26]

$$G_{c(RT)} = 6w(\lambda_c)a/\sqrt{\lambda_c}, \tag{1}$$

where $\lambda_c$ is the critical stretching ratio at fracture of a precut specimen. This formulation derives from the Griffith style energy balance considerations. However, it was found since the beginning[8] that $G_{c(RT)}$ is orders of magnitude higher than surface fracture energy. This discrepancy has led researchers in the community to seek dissipative mechanisms[23] to account for the observed much greater energy release and to realize that fracture at a finite speed cannot be idealized as involving only one monolayer of debonding, which is the case treated by the Lake and Thomas model[27] to describe the threshold for fatigue failure.

By directly examining the stress buildup at crack tip in a precut PBR specimen we explore the merit to adopt the Westergaard-Irwin's local stress approach[10, 11] that recognizes $K=\sigma\sqrt{\pi a}$ as the loading parameter. Given the well-established stress optical relationship (SOR), shown in Fig.

1*A*, i.e., $\Delta n = C\sigma$, with $C = 2.92$ GPa$^{-1}$, we apply *str*-POM to quantify the stress field as a distance from the cut tip within a spatial resolution better than 10 μm. At $\theta = \pi/3$ to the crack propagation direction, the tensile stress field $T(r) = (\sigma_{yy} - \sigma_{xx})$ is also the principal stress difference,[16] thus linearly related to the birefringence $\Delta n$, i.e., $T = \Delta n(r)/C$, where C is the same stress-optical coefficient C, measured in Fig. 1*A*.

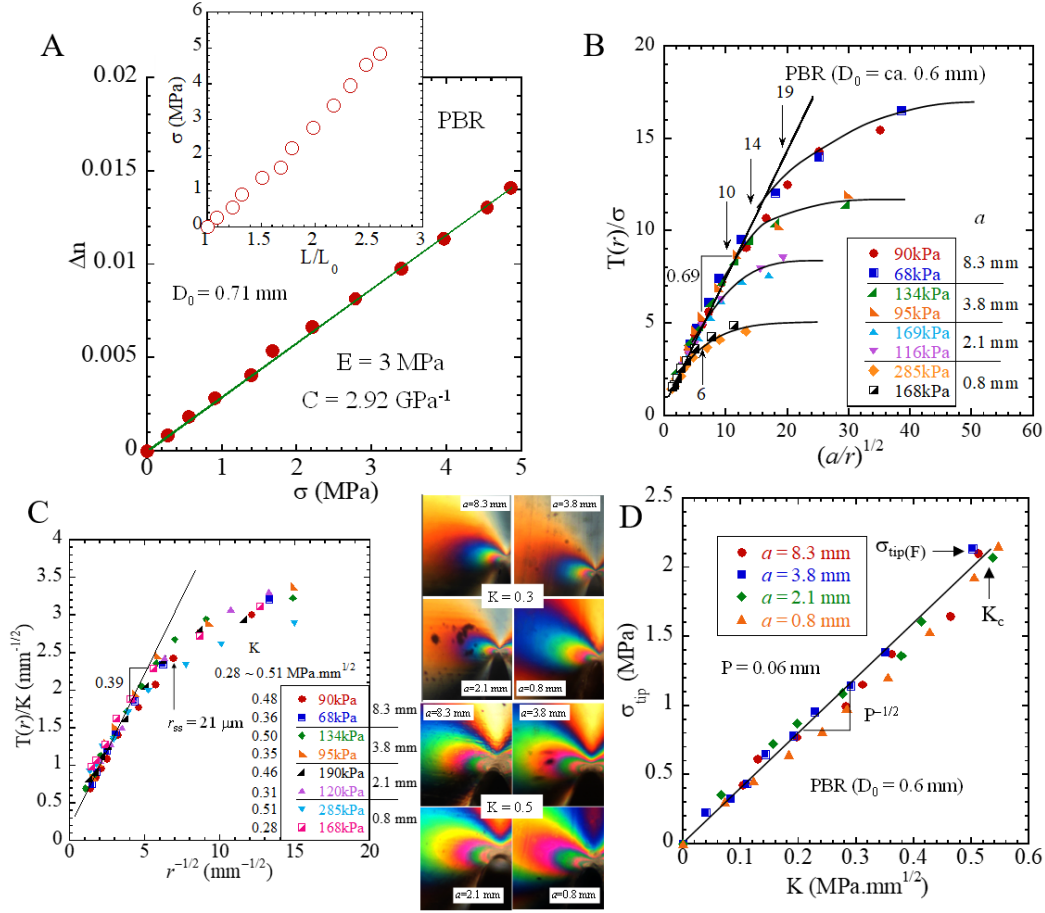

**Fig. 1** (*A*) Stress optical relationship (SOR) through simultaneous mechanical and birefringence measurements of uncut PBR specimen at V/L = 0.05 min$^{-1}$. The inset is the stress vs. strain curve of the specimen in terms of true stress σ and stretching ratio $\lambda = L/L_0$. (B) Normalized tensile stress field T/σ, read from SOR in (A) as a function $(a/r)^{1/2}$. Eq. (2), represented by the straight line, holds up to various values, i.e., 6, 10, 14 and 19 of $(a/r_{ss})^{1/2}$. For $r < r_{ss} = 21$ μm, T/σ saturates toward finite values. The curves are drawn to indicate the trend of stress saturation. (C) Various curves in (B) collapse when K instead of nominal load σ is used to normalize T, revealing a common $r_{ss}$, independent of K that ranges from 0.28 to 0.51 MPa.mm$^{1/2}$. Two groups of images show the same birefringence level at the cut tip at K = 0.3 and 0.5 MPa.mm$^{1/2}$. The different stages of stretching in (B) and (C) are labeled by the values of engineering stress. (D) Master curves of $\sigma_{tip}$ vs. K for the four values of cut size *a*, revealing a linear scaling law, characterized by a new length

scale P. The last points are the values of $K_c \sim 0.5$ MPa.mm$^{1/2}$ and $\sigma_{F(inh)} \sim 2$ MPa at the onset of fracture.

Quantitative spatial mapping of $\Delta n$ shows in Fig. 1B that the stress field naturally saturates at the cut tip for four different cut sizes at various stages of tensile extension of the precut specimen, which is described in Supporting Information (SI.A). In such a plot, a Westergaard like solution describes the straight line in Fig. 1B, revealing

$$T = \sigma_{yy} - \sigma_{xx} = 0.69\sigma(a/r)^{1/2} \text{ for } r > r_{ss}. \tag{2}$$

But two features deviate from this textbook description of LEFM. First, instead of stress divergence, described by Eq. (2), stress saturation is observed under all eight conditions. This stress saturation zone (SSZ) emerges at a common distance $r_{ss}$ from the tip, independent of cut size and load level, as indicated by values of 6, 10, 14 and 19 for $(a/r_{ss})^{1/2}$, corresponding to the respective values of $a$, so that $r_{ss} = 20$-$22$ μm under all conditions. To emphasize that the size of the SSZ is constant, i.e., unchanging with K, we replot Fig. 1B to obtain Fig. 1C. For the same $r_{ss}$, K varies from 0.28 to 0.51 MPa. Textbooks[4, 5, 12] on LEFM deal with the Westergaard stress singularity by proposing Irwin plastic zone (also known as damage or process or messy zone) involving a fixed yield stress $\sigma_{YS}$, with the zone size growing quadratically with K. The SSZ is clearly not a process zone, having a fixed size instead of varying by a factor of $(0.51/0.28)^2 = 3.3$. Second, instead of locking onto a hypothetical $\sigma_{YS}$, the tip stress $\sigma_{tip} = T(r \rightarrow 0)$ increases linearly with K until the point of fracture. Consequently, the stress field T normalized by K collapses in Fig. 1C, apart from a small correction by the far-field load that manifests as a non-zero intercept on the ordinate.

The linear scaling relationship in Fig. 1D between $\sigma_{tip}$ and K reveals[16] a characteristic length scale P that is constant at various stages of external loading including the onset of fracture, characterized by ($K_c$, $\sigma_{tip(F)}$). Thus, in terms of P, the critical stress intensity factor $K_c$ acquires a new expression, besides its operational definition, given by

$$K_c = \sigma_c\sqrt{\pi a} = \sigma_{F(inh)}\sqrt{P}, \tag{3}$$

where we suggest the tip stress at fracture as revealing inherent strength, i.e., $\sigma_{tip(F)} = \sigma_{F(inh)}$. Equation (3) explains why $K_c$ is material specific: Both $\sigma_{F(inh)}$ and P are material specific. Emergence of P originates from the existence of SSZ, which also persists during frack propagation as shown in Supporting Information (SI.A).

**Table 1 Toughness $K_c$ and $G_c$ of PBR for different cut sizes†**

| cut size $a$ | thickness $D_0$ | $\varepsilon_c$ | $\sigma_c$ | $K_c = \sigma_c\sqrt{\pi a}$ | $G_c = K_c^2/E$ | $w(\lambda_c)$ | $G_{c(RT)}$ |
|---|---|---|---|---|---|---|---|
| 8.4 | 0.62 | 0.037 | 0.100 | 0.51 | 87 | 1.8 | 89 |
| 3.8 | 0.61 | 0.059 | 0.148 | 0.51 | 87 | 4.4 | 97 |
| 2.1 | 0.55 | 0.075 | 0.209 | 0.54 | 105 | 7.1 | 95 |
| 0.8 | 0.61 | 0.137 | 0.347 | 0.55 | 101 | 23.0 | 100 |

† : Lengths $a$ and $D_0$ are in the unit of millimeters, $\varepsilon_c = \lambda_c - 1$, stress $\sigma_c$ and modulus E in the unit of MPa, $G_c$ and $w_c$ are in the units of J/m$^2$ and kJ/m$^3$ respectively. Note that if we adopt Eq. (12) in Section III for $K_c$, $G_c$ would be slightly higher than $G_{c(RT)}$.

Moreover, it is worth indicating that for the present PBR samples the fracture toughness can be computed either from the operational definition of $K_c$ for SEN or the Rivlin-Thomas formula Eq. (1). Table 1 shows the comparison between the two different estimates of $G_c$. They are clearly comparable.

*II.B Nature of fracture in pure shear*

Rivlin and Thomas proposed an elegant protocol to account for the energy release during fracture of rubber. Specifically, the critical energy release rate was argued to show a simple form of $G_{ps(c)} = w_c h_0$ in the Rivlin-Thomas scenario, as shown in Supporting Information (SI.C). Since $G_{ps(c)}$ is usually constant, the Rivlin-Thomas expression explicitly suggests that a taller specimen with larger $h_0$ would undergo fracture at smaller strain because it has a larger volume and can store the same amount of energy at a lower energy density, i.e., $w_c = w(\lambda_c) \sim 1/h_0$: Larger $h_0$ corresponds to smaller $\lambda_c$. However, the results from the preceding II.A prompt us to look for a different interpretation of the Rivlin-Thomas formulation.

We proceed to examine three EPDM specimens of $h_0 = 18$, 48 and 120 mm in pure shear, where precut length $c$ is half of the specimen width W of 100 mm. For pure shear, we adopt a different notation to denote cut length as $c$ in place of $a$ for SEN. The finding of preceding section informs us that fracture in pure shear would occur when the tip stress grows to reach $\sigma_{F(inh)}$. The characterization of pure shear fracture begins with experiments to determine the SOR for the EPDM specimens, from which the local stress field can be mapped out. Fig. 2*A* shows that the SOR is roughly a linear relation between tensile stress and birefringence. The slight deviation from the linearity might be due to onset of non-Gaussian chain stretching. We show in Supporting Information (SI.B) that SOR can actually be applied to show a difference in the stress states between uniaxial extension and pure shear.

Fig. 2*B* presents the relations between nominal load and nominal strain for both the uncut and precut specimens. Inspired by Fig. 1*B*, we make a similar observation as shown in Fig. 2*C* and reveal a similar feature that the local stress ceases to increases as $r^{-1/2}$ for $r < r_{ss} = 0.055$ mm, where the stress field T($r$) at $\theta = \pi/3$ is estimated using the stress-optical relationship from Fig. 2*A*. In place of the variable $a$, we have the specimen height $h_0$, which is the only pertinent length scale in the configuration.[2] According to the theoretical analysis in Supporting Information (SI.C), for linear elastic materials such as the present PBR and EDPM samples with cut length $c$ equal to *half* of specimen width W, we have the following expression for the load parameter $K_{ps}$

$$K_{ps} = \sigma\sqrt{h_0} \tag{4}$$

in place of $K_{SEN} = \sigma\sqrt{\pi a}$. Like the Westergaard loading parameter $K_{SEN}$ for SEN, we expect the pure shear to show stress intensification through $K_{ps}$ as T($r$) ~ $K_{ps}/(2\pi r)^{1/2}$ at cut tip. Indeed Fig. 2*C* shows that away from the tip, T($r$) scales as

$$T = 0.45\sigma(h_0/r)^{1/2} \text{ for } r > r_{ss}, \tag{5}$$

which is given by the straight line. In other words, all data involving three different values of $h_0$ fall on the same straight line until the emergence of SSZ in the same distance $r$ from the cut tip, at $(h_0/r_{ss})^{1/2}$ = 18, 29 and 50, i.e., at $r_{ss}$ = 0.055 mm. This SSZ's size is independent of $K_{ps}$. Thus the SSZ has little to do with Irwin's plastic zone concept.

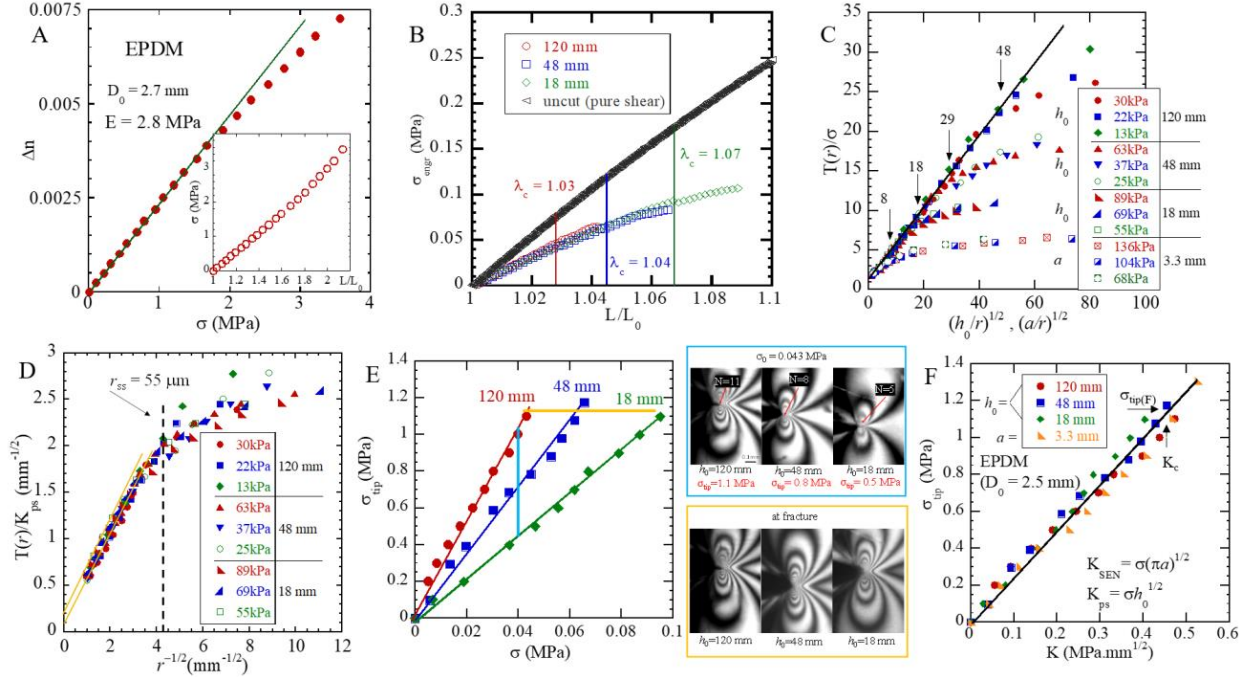

**Fig. 2** (*A*) Stress optical relationship (SOR) through simultaneous mechanical and birefringence measurements of uncut EPDM specimen at V/L = 0.05 min$^{-1}$. The inset is the stress vs. strain curve of the specimen in terms of true stress $\sigma$ and stretching ratio $\lambda$ = L/L$_0$. (B) Engineering stress vs. nominal strain $\lambda$ = L/L$_0$ of uncut ($h_0$, W = 100 mm) and precut (cut length $c$ = W/2 = 50 mm) specimens of different heights $h_0$ = 120, 48 and 18 mm. (C) Normalized tensile stress field T/$\sigma$, at different values of $\sigma$ read from SOR in (A) as a function $(h_0/r)^{1/2}$ or $(a/r)^{1/2}$, where the last group of data is a case of SEN with $a$ = 3.3 mm. Eq. (5) represented by the straight line holds up to various values of $(a/r_{ss})^{1/2}$, i.e., 8, 18, 29 and 48 of $(h_0/r_{ss})^{1/2}$. For $r < r_{ss}$ = 55 μm, T/$\sigma$ saturates toward finite values. Here and in the subsequent (D), we label different stages of stretching with nominal stress, i.e., engineering stress. (D) Various curves in (C) collapse when $K_{ps}$ of Eq. (4) is used to normalize T, revealing a common $r_{ss}$, independent of $K_{ps}$. The different stages of stretching in (C) and (D) are labeled by the values of engineering stress. (E) Tip stress ($\sigma_{tip}$) equal to T($r \to 0$) as a function of nominal load $\sigma$ for $h_0$ = 120, 48 and 18 mm. The upper group of images depict the birefringence field near cut tip at a common load, indicated by the vertical (light blue, color online) line at 0.043 MPa; the lower group of images show the birefringence field at onset of fracture involving a common level of $\sigma_{tip(F)}$ around 1.14 MPa, indicated by the horizontal orange line. (F) Master curves of $\sigma_{tip}$ vs. K for the three values

of $h_0$, revealing a linear scaling law, involving a characteristic length scale P, the triangles denote the data from SEN.

To emphasize the importance of the loading parameter $K_{ps}$ whose operational definition is given in Eq. (4), similar to Fig. 1*C*, we can make Fig. 2*D* by normalizing the local tensile stress T with $K_{ps}$ to illustrate analogous features. In the scaling regime, i.e., for $r > r_{ss}$, like Fig. 1*C*, the data show the same slope. However, they are not expected to collapse onto a single straight line because these lines have different intercepts given by $h_0^{-1/2}$ in Fig. 2*D* and by $a^{-1/2}$ in Fig. 1*C*.

The existence of SSZ allows the tip stress $\sigma_{tip}$ to be measured as T($r=0$) and expressed for all three values of $h_0$ as a function of load $\sigma$. Specifically, Fig. 2*E* shows how the tip stress varies with $h_0$ at various nominal loads. For example, at a common load of 0.043 MPa, indicated by the vertical line, $\sigma_{tip}$ is higher for higher $h_0$. Reading Fig. 2*E* "horizontally", we see the same tip stress is reached at a lower load for higher $h_0$. The loading parameter $K_{ps}$ of Eq. (4) can collapse the three lines in Fig. 2*E*, as shown by Fig. 2*F*.

Similar to Fig. 1*D*, Fig. 2*F* discloses a similar scaling law: $\sigma_{tip} = K_{ps}/\sqrt{P}$, obeyed by both pure shear and SEN data. The overlapping of the triangles with other three sets of data confirm that Eq. (4) is the correct expression for $K_{ps}$. Fracture occurs in both SEN and pure shear at a comparable tip stress that is bounded by the inherent strength, with $K_{SEN(c)} \approx K_{ps(c)}$, which implies equivalently $G_{SEN(c)} \approx G_{ps(c)}$. In other words, similar to Eq. (3), we have

$$K_{ps(c)} = \sigma_c\sqrt{h_0} = \sigma_{F(inh)}\sqrt{P} \tag{6}$$

Like $K_c$, $\sigma_c$ acquires its meaning through either Eq. (6) for either pure shear or Eq. (3) for SEN: it is given by $\sigma_{F(inh)}$ and the ratio of either $h_0$/P or $a$/P. Unlike $K_c$, $\sigma_c$ is not a material constant because it depends on experimental parameters such as $h_0$ and $a$.

**Table 2 Toughness $K_c$ and $G_c$ of EPDM: pure shear vs. SEN†**

| $h_0$ or $a$ | thickness $D_0$ | $\varepsilon_c$ | $\sigma_c$ | $K_c$ | $G_c = K_c^2/E$ | $w(\lambda_c)$ | $G_{c(RT)}$ |
|---|---|---|---|---|---|---|---|
| 120 | 2.50 | 0.026 | 0.043 | 0.47 | 80 | 1.1 | 66 |
| 48 | 2.56 | 0.043 | 0.066 | 0.46 | 74 | 2.7 | 65 |
| 18 | 2.51 | 0.064 | 0.095 | 0.40 | 58 | 6.1 | 55 |
| 3.3 | 2.49 | 0.075 | 0.163 | 0.53 | 99 | 5.7 | 106 |

† : Lengths $h_0$, $a$ and $D_0$ are in the unit of millimeters, $\varepsilon_c = \lambda_c - 1$, stress $\sigma_c$ and modulus E in the unit of MPa, $G_c$ and $w$ are in the units of J/m² and kJ/m³ respectively. $K_c$ is evaluated from Eq. (4) for $K_{ps}$ for the first three rows. $K_c$ is the bottom row follows from $K_{SEN(c)} = \sigma_c\sqrt{\pi a}$ in MPa·mm$^{1/2}$. $G_{c(RT)}$ is evaluated according to either Eq. (13) in Section III.C for pure shear or Eq. (1) for SEN.

Since Fig. 2*C* reveals $r_{ss}$ to be comparable in both pure shear and SEN, we expect the slope of the $\sigma_{tip}$ vs. K to be comparable, i.e., involving a common P, leading to a collapse of the SEN data onto the master curve in Fig. 2*F*. Clearly, Eq. (4) is the correct formula for $K_{ps}$ at $c = W/2$. On the other hand, $K_c$ seems slightly higher in SEN than in pure shear. This leads to higher $G_c$ for SEN as shown in Table 2 in comparison with pure shear. Also listed in Table 2 is $G_{c(RT)}$ given by either Eq. (1) for SEN and Eq. (13) for pure shear. The discrepancy of $G_c$ between pure shear and SEN seem to arise from a difference in $\sigma_{F(inh)}$ that varies from SEN to pure shear. We return to this point in Section III.A. Moreover, the relation between SEN and pure shear is further explored in Supporting Information (SI.D) based on a second rubber, i.e., PBR.

*II.C Thickness dependence of toughness $K_c$: local characterization*

II.C.1 Precut

If polymer fracture is dictated by crack tip reaching inherent strength, it is natural to ask whether and how inherent strength varies with the mode of deformation. Since the state of stress is known to change from plane stress to plane strain upon thickness increase, using PBR with SEN, it is straightforward to find out how the critical condition for fracture depends on specimen thickness. Making precut of comparable size $a$ around ca. 4.0 mm, Fig. 3*A* indicates that fracture occurs at a comparable $K_c$ because the crack propagation occurs at a similar load level of $K_c = 0.49$ MPa.mm$^{1/2}$, involving $\sigma_{engr} = 0.124$ MPa for $D_0 = 5.16$ mm and $a = 4.4$ mm and $\sigma_{engr} = 0.13$ MPa for $D_0 = 1.33$ mm and $a = 4.0$ mm as well as $\sigma_{engr} = 0.14$ MPa for $D_0 = 0.61$ and $a = 4.0$ mm.

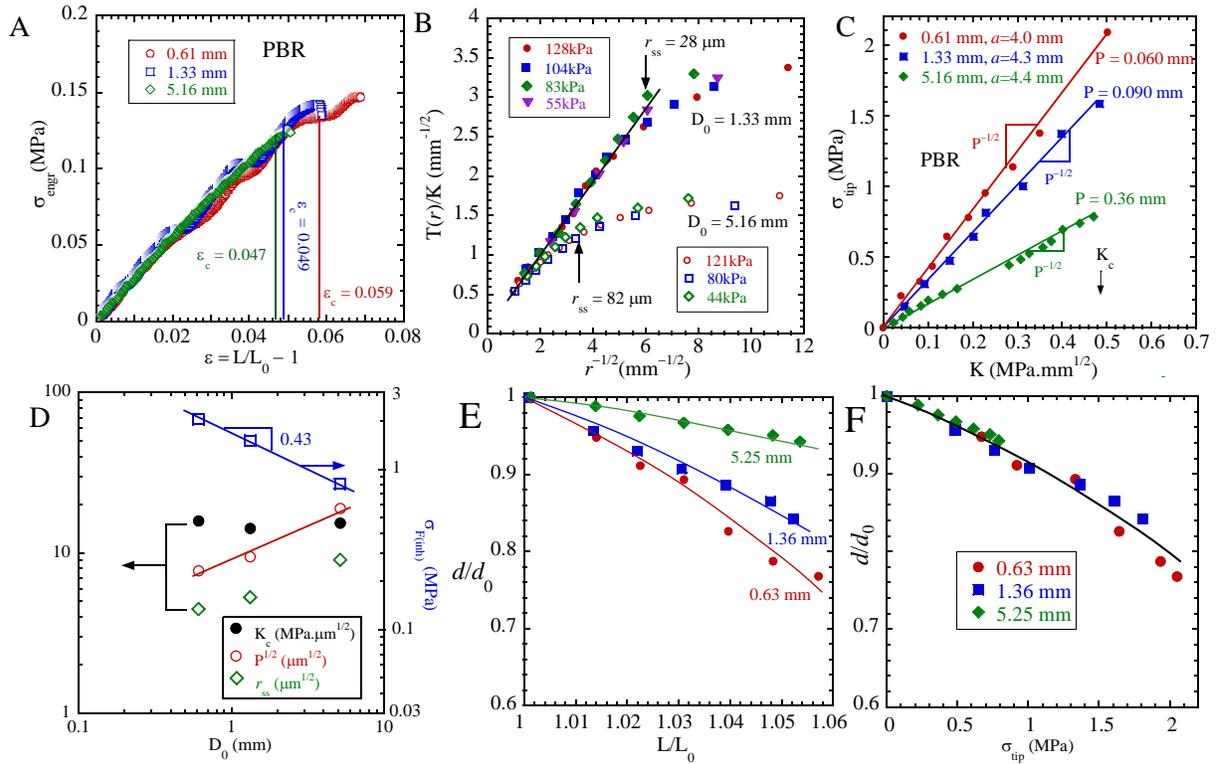

**Fig. 3** (*A*) Engineering stress vs. nominal strain $\varepsilon = L/L_0 - 1$ of precut PBR specimens of different $D_0 = 0.61$, 1.33 and 5.16 mm. (*B*) Normalized tensile stress field $T/K$ at various nominal load $\sigma = 55, 83, 104$ and 128 kPa as a function of $r^{-1/2}$ for $D_0 = 1.33$ and 5.16 mm. (*C*) Tip stress $\sigma_{tip}$ vs. K, showing different slopes and thus revealing different P values at different values of $D_0$ yet a common $K_c$. (*D*) SSZ size $r_{ss}$ and corresponding P as well as $\sigma_{F(inh)}$ as a function of specimen thickness $D_0$. The opposing scaling of P and $\sigma_{F(inh)}$ leads to independence of $K_c$ on $D_0$, denoted by filled circles. (*E*) Local thickness decrease at cut

tip during extension, i.e., as a function of nominal draw ratio $L/L_0$. (F) Master curve of $d/d_0$ against the tip stress $\sigma_{tip}$, with the smooth curve as guide for the eye.

We apply *str*-POM observation of the cut tip to show how $K_c$ reaches a common level independent of thickness $D_0$. The data in Fig. 3*B* are unexpected, revealing not only emergence of SSZ, analogous to data in Fig. 1*C* and 2*D*, but also dependence of SSZ size $r_{ss}$ on the specimen thickness $D_0$. Fig. 3*B* also suggests that at comparable loads (and a common nominal strain), e.g., common values of K, the tip stress $\sigma_{tip}$ changes with $D_0$, lower for a larger $D_0$, as show in Fig. 3*C*, which also reveals three separate values for P, consistent with the fact that $r_{ss}$ depends on $D_0$ (cf. Fig. 3*B*). The thickness effect can be more explicitly described in Fig. 3*D*: $\sigma_{tip(F)} = \sigma_{F(inh)}$ scales with $D_0$ to a negative power of 0.43 while $P^{1/2}$ roughly increases with $D_0$ in a power law with a positive exponent of 0.43 so that $K_c = \sigma_{F(inh)} P^{1/2}$ stays constant at 0.49 MPa.mm$^{1/2}$.

We reiterate the explicit message revealed in Fig. 3*B*: At all values of K, the thicker specimen with $D_0 = 5.3$ avoids stress buildup by stress saturation at a greater distance from the crack tip. To further characterize the thickness effect, local thickness at the cup tip is measured during stretching in separate tests. Fig. 3*E* shows that the local thickness at the tip decreases with nominal strain $L/L_0$, going faster for the thinner specimen. The greater thickness change at the tip amounts to greater stress buildup. In other words, the thicker specimen has greater capacity to avoid the buildup at the tip. Fig. 3*F* indicates that the higher tip stress is actually associated with the greater thickness change at the tip.

II.C.2 Preload

As stated in the beginning of Section II.C.1, catastrophic fracture occurs at comparable nominal loads or common $K_c$ as indicated by Fig. 3*A*, Fig. 3*C* and Fig. 3*D*. To confirm that the critical condition for fracture is indeed independent of thickness, we pre-load uncut PBR specimens to different levels and then force a blade from one edge into the specimen to produce a crack of increasing length until catastrophic fracture at a critical cut size $a_c$. Specifically, we discretely preload PBR specimens to three levels of $\sigma_{engr} = 0.12$, 0.15 and 0.20 MPa for each of the three thicknesses using a constant load mode. As shown in Table 3, for three preloads of 0.2, 0.15 and 0.12 MPa, the blade advances to three different sets of distances before spontaneous fracture, revealing a similar cut size for a given preload, roughly independent of $D_0$. Here the values of $a_c$ can be readily measured from the fractured specimens since the catastrophic fracture produces much rougher surfaces relative to that made by the blade, as shown by images of fracture surface in Supporting Information (SI.E). The duration of blade advancement is characterized by Instron measurement of clamp movements, also presented in Supporting Information (SI.E). With increasing load $\sigma$, $a_c$ decreases, leading to a constant $K_c = \sigma\sqrt{\pi a_c}$. Specifically, Table 3 lists the values of $K_c$ that are roughly thickness independent, showing slight increase with the load $\sigma$, ranging from 0.71 MPa.mm$^{1/2}$ for $\sigma_{engr} = 0.2$ MPa to 0.56 MPa.mm$^{1/2}$ for $\sigma_{engr} = 0.12$ MPa, which is comparable to the level obtained from precut specimens (cf. Table 1). In other words, these preload tests confirm that the toughness is indeed constant independent of the specimen thickness.

**Table 3  Fracture of PBR under preload**

| 0.2 MPa | | | |
|---|---|---|---|
| $D_0$ (mm) | 0.68 | 1.37 | 5.40 |
| $a_c$ (mm) | 2.87 | 3.48 | 2.97 |
| $K_c$ (MPa.mm$^{1/2}$) | 0.65 | 0.71 | 0.67 |
| **0.15 MPa** | | | |
| $D_0$ (mm) | 0.53 | 1.43 | 5.27 |
| $a_c$ (mm) | 4.92 | 5.17 | 5.23 |
| $K_c$ (MPa.mm$^{1/2}$) | 0.62 | 0.64 | 0.65 |
| **0.12 MPa** | | | |
| $D_0$ (mm) | 0.63 | 1.44 | 5.31 |
| $a_c$ (mm) | 5.92 | 6.37 | 6.71 |
| $K_c$ (MPa.mm$^{1/2}$) | 0.54 | 0.56 | 0.58 |

## III. Discussion

*III.A Inherent strength $\sigma_{F(inh)}$*

We have defined the tip stress at fracture as inherent strength $\sigma_{F(inh)}$. The *str*-POM observations in Section II.C.1 indicate that $\sigma_{F(inh)}$ is not constant and varies with specimen thickness. The origin of this variation remains elusive. On the other hand, it is clear that $\sigma_{F(inh)}$ is appreciably lower than the fracture strength (i.e., breaking stress) of uncut specimen, denoted by $\sigma_b$. To understand this difference, it is helpful to note that the uniaxial extension (along Y axis) is described by a single non-zero component $\sigma_{yy}$ with $\sigma_{xx} = \sigma_{zz} = 0$, which can be quantified through SOR as

$$\sigma_{yy} = \sigma_{yy} - \sigma_{xx} = \Delta\sigma = \Delta n/C. \tag{7}$$

In contrast, according to the Westergaard description of the stress field near the tip, at $\theta = \pi/3$, $\sigma_{yy} = 3\sigma_{xx}$ so that the SOR amounts to

$$\sigma_{yy} = (3/2)\Delta n/C, \tag{8a}$$

$$\sigma_{xx} = \sigma_{yy}/3 = (1/2)\Delta n/C, \tag{8b}$$

and

$$\sigma_{tip} = T(r \to 0, \theta = \pi/3) = \Delta n(r \to 0)/C, \tag{9}$$

which is lower than $\sigma_{yy}$ by a factor of 2/3. We expect *str*-POM observations to reveal

$$(3/2)\sigma_{F(inh)} < \sigma_b \tag{10}$$

because fracture occurs at $\sigma_b$ under unconstrained condition of $\sigma_{xx} = \sigma_{zz} = 0$ while fracture at the tip occurs under severe constraint of $\sigma_{xx} = \sigma_{yy}/3$ and $\sigma_{zz}$ ranging from 0 to $2\sigma_{xx}$ (with Poisson ratio $v = \frac{1}{2}$). To clarify, Eq. (10) stems from our assertion that elastomeric fracture under constraint (i.e., with $\sigma_{xx} > 0$ and $\sigma_{zz} \geq 0$) would occur at lower $\sigma_{yy}$ than $\sigma_b$ measured from the unconstrained uniaxial extension of uncut specimen. While a chain-level theory of inherent strength for elastomers under different deformation modes is non-existent, the preceding assertion is self-evident.

It is necessary to further recognize that the value of $\sigma_{F(inh)}$ (relative to $\sigma_b$, which we regard as inherent strength for uniaxial stretching) may also depend on whether $\sigma_{zz}$ is zero or finite. For example, apparently, the specimen thickness regulates the stress state at cut tip, causing $\sigma_{F(inh)}$ to vary, as implied by the data in Section II.C.1. Moreover, the difference found in $K_c$ between SEN and pure shear, as listed in Table 2, may also indicate a notable difference in $\sigma_{F(inh)}$. It is plausible that the mode of global extension, e.g., uniaxial vs. pure shear, can influence the state of deformation at the cut tip, leading to differences in $\sigma_{F(inh)}$.

### III.B Single-edge notch (SEN)

For rubber such as the present PBR and EPDM, the stress buildup is found to saturate so that it is feasible to characterize the tip stress $\sigma_{tip}$ as a function of the nominal load $\sigma$ or stress intensity factor K based on its operational definition. The stress buildup at the cut tip is prescribed by the combination of nominal load $\sigma$ and cut size *a*, as anticipated by linear elastic fracture mechanics. Indeed, fracture from precut specimen is a local event, dependent on the stress intensification at the cut tip and independent of global configurations such as the variation from SEN to pure shear. For SEN, the tensile stress field T at a distance *r* from the tip and angle $\theta = \pi/3$, after normalization by the nominal load $\sigma$, can be expected,[16] according to the Westergaard solution,[10] as

$$T/\sigma = 1.12\sin(\pi/3)(K_{exp}/K_{theor})\sqrt{a/2r} + 1, \tag{11}$$

with the theoretical operational definition[4]

$$K_{theor} = 1.12\sigma\sqrt{\pi a}. \tag{12}$$

Unity in Eq. (10) denotes the fact that the far-field stress is comparable to the nominal load. Comparing Eq. (11) with the experimental result described by Eq. (2), we have $K_{exp} = K_{theor}$.

### III.C Pure shear

For pure shear, analogous to Eq. (1), Rivlin and Thomas proposed the following widely used expression for $G_{ps(c)}$ as

$$G_{ps(c)} = w(\lambda_c)h_0/2, \tag{13}$$

which is re-drived in Supporting Information (SI.C). Parallel to the case of SEN, we can envision stress intensification in pure shear to be characterized by a stress intensity factor $K_{ps}$, involving an expression similar to Eq. (12). The exact form of $K_{ps}$ given in Eq. (4) is derived in the Supporting Information (SI.C). Specifically, because of the scaling form in Eq. (4), the following expression can be expected in the $K_{ps}$ annulus for the local stress field driven by the nominal load $\sigma$

$$T/\sigma = B\sqrt{h_0/r} + 2, \text{ at } \theta = \pi/3, \tag{14}$$

where the factor of two arises from the special case of the cut length $c$ in pure shear equal to half width, W/2, so that the nominal load $\sigma$ is half of the far-field stress $\sigma_{uc}$ equal to that of uncut specimen. Eq. (14) suggests that the pure shear data should be analyzed by plotting T/$\sigma$ against normalized distance $\sqrt{h_0/r}$ for different values of $h_0$. This is indeed the case according to the experimental data in Fig. 2C, partically described by Eq. (5), showing $B = 0.45$.

Rivling-Thomas[8] varied $h_0$ by a factor of three to validate Eq. (13) that $G_c$ is a material constant, independent of $h_0$, i.e., $w_c \sim 1/h_0$. The form of Eq. (13) conveys the message that energy balance argument is valid: The taller specimen has more volume to store energy and thus only needs to involve a lower level of strain energy to produce the same energy release rate. In their subsequent papers from II to XI, Rivlin and Thomas did not return to the question of whether and why Eq. (13) holds for different values of $h_0$. Soon after their Paper I Thomas became aware[28] that $G_c$ may be related to the strain distribution around a blunted crack tip. Unfortunately, Thomas did not proceed to show that Eq. (13) may have its origin in the local stress intensification that increases with $h_0$, as shown by Eq. (4) and Eq. (14): At a given nominal strain $\lambda$, the local stress field near crack tip is higher at a larger $h_0$.

*Str*-POM observation of the stress field near crack tip unravels the hidden meaning of Eq. (13). In other words, we explain for the first time why $w_c \sim 1/h_0$, as implied by Eq. (13). We show by the specific experiments elaborated in Section II.B that $K_{ps}$ of Eq. (4) indeed holds, i.e., the tip stress actually builds up in linear proportion to $K_{ps} \sim \sqrt{h_0}$, or more generally Eq. (14) holds true. To our knowledge, there is no theoretical derivation of Eq. (14) for linear elastic materials in pure shear configuration. We have established this relation by experiment and elucidates the actual meaning of the Rivlin-Thomas formula, Eq. (13). It is a stress criterion that is behind Eq. (13). G → $G_{ps(c)}$ is the consequence of $\sigma_{tip}$ → $\sigma_{F(inh)}$, which is universal, e.g., is also the fracture criterion for SEN configuration. Using Eq. (6), we can arrive at an expression for toughness $G_c$ as

$$G_c = K_c^2/E = (\sigma_c^2/E)h_0 = 2w_F P, \text{ with } w_F = (\sigma_{F(inh)})^2/2E. \tag{15}$$

We note that the same expression applies to SEN. Thus, Eq. (3) and Eq. (15) describe the actual meaning of toughness $K_c$ and $G_c$, with $G_c$ computable from $K_c$ while the reverse does not hold – we cannot derive stress intensification characterized by $K_c$ from $G_c$.

## IV. Conclusion

Spatial-temporal resolved POM (*str*-POM) observations show that the stress buildup at crack tip saturates at all levels of applied load at a sizable distance (beyond 20 μm) from crack tip. Existence of a stress saturation zone (SSZ) permits tip stress $\sigma_{tip}$ to be determined as a function of nominal load or stress intensification factor K (through its operational definition). The observed linear growth of $\sigma_{tip}$ with K also contrasts sharply with basic notion in LEFM that stress intensification at crack tip ought to approach a common yield stress at all load levels. Moreover, this linear scaling reveals a new characteristic length scale $P = (K/\sigma_{tip})^2$. Since the size of SSZ is independent of K, SSZ cannot regarded as an Irwin process zone. On the other hand, sizable process zone has been reported for toughened elastomers.[29] Separately, theoretical studies have suggested[30-33] that the local stress can saturate upon approaching the tip due to the finite curvature of crack tip.

Based on three independent sets of *str*-POM studies, we show that elastomeric fracture may be characterized in terms of a stress-based criterion rather than Griffith energy-balance argument. Independent of geometric configuration, e.g., for different cut sizes in single-edge notch, different specimen thicknesses, pure shear (planar extension) vs. uniaxial extension, the onset of fracture is shown to be governed by the criterion of tip stress reaching a threshold level, designated as inherent strength. This stress criterion allows us to explain how the LEFM phenomenology emerges in reality. For example, Eq. (3) shows (a) why $K_c$ is a material constant – because both inherent strength $\sigma_{F(inh)}$ and P are material specific, and (b) what determines the magnitude of $K_c$ and $G_c = K_c^2/E$, per Eq. (3) and Eq. (15). In other words, the *str*-POM measurements have allowed us to uncover the hidden variable in LEFM, i.e., P whose origin is the emergence of SSZ; the toughness $K_c$ or $G_c$ is determined by two parameters $\sigma_{F(inh)}$ and P.

The *str*-POM observation of stress buildup at crack tip suggests that the widely used pure shear protocol of Rivlin and Thomas[8] to characterize toughness of elastomers has its origin in the tip stress scaling with specimen height $h_0$ as $\sigma_{tip} \sim K_{ps} = \sigma\sqrt{h_0}$. In other words, we have uncovered the true meaning of Rivlin-Thomas Eq. (13) that stress intensification depends on $h_0$ and fracture is tip stress controlled.

Finally, through *str*-POM, we are now able to show how elastomeric fracture occurs at a common $K_c$, independent of specimen thickness $D_0$,[18] unlike the case of plastics.[17] Specifically, the local deformation at crack tip varies with $D_0$, showing stress saturation at a larger distance from the tip for a thicker specimen, corresponding to weaker deformation at the tip, with fracture observed to commence at a lower tip stress, i.e., lower inherent strength.

**Materials and Methods**
Materials and Sample Preparation

Polybutadiene rubber (PBR) from The Goodyear Tire & Rubber Company (BUD1207) and ethylene propylene diene monomer (EPDM) from Lion Elastomers (Royalene 511) were studied in this work. Only dicumyl peroxide is added as the crosslinking agent to guarantee transparency in PBR. PBR sheets with thickness 0.6-5.3mm were cured by heat compression molding with 1 phr dicumyl peroxide (Acros Organics) at 160 °C for 60 minutes. EPDM sheets with thickness 2-2.5 mm were prepared by Gregory Brust at Lion Elastomers using tri-functional crosslinker SR-350 (1~3 phr), and peroxide DiCup R (3~4 phr), at 170 °C for 20 minutes.

Dogbones were cut using type V die (ASTM D638). They are used to establish the stress-optical relationship. Strip-shaped specimens were cut into desired dimension by a paper cutter. Precut specimens involving edge notch, i.e., either single-edge notch (SEN) or pure shear with cut length being half of the specimen width, are made by gently pushing a razor blade (Feather Safety Razor Co., Ltd.) into specimens. The cut size to width ratio (*a*/W) for all SEN is kept smaller than 0.2. For cut sizes smaller than ca. 4 mm, the length and width of the specimen is chosen to be 50×20 mm²; for cut sizes larger than 4mm, length and width is chosen to be 80×40 mm².

Methods

Tensile experiments were carried out on an Instron 5543 tensile tester at room temperature. Photoelastic characteristics of PBR and EPDM were mapped both spatially and temporally around the cut tip using a *str*-POM setup[16] described in a previous study. To capture the development of birefringence during stretching, video camera (Mokose C100) with 4K resolution was used along with two different magnification-adjustable micro-lenses (Edmund Industrial Optics (EIO) and Hayear model HY-180XA). The *str*-POM observations in Section II.A involved 8× magnification on EIO. For data in Section II.B, Hayear lens at 2.25× was used. Finally, for *str*-POM observations presented in Section II.C, magnification on EIO was set at 7×, 5× and 2.5×, corresponding to $D_0$ = 0.61, 1.33 and 5.16 mm respectively.

The monochromatic light source for most of our *str*-POM observations is a low-pressure sodium lamp (Phillips). For samples with thickness less than 0.7mm, we adopt white light and use Michel-Levy charter to make more accurate measurements of birefringence. For preload tests presented in II.C.2, the same razor blade is advanced into one of the two edges of a strip-shaped specimen until fracture.

**CRediT authorship contribution statement**
Zehao Fan: Development of quantitative birefringence measurements, Data curation, Formal analysis, Validation, Writing – Materials and Methods section. Shi-Qing Wang: Supervision, Conceptualization, Methodology, Formal analysis, Validation, Writing – original draft, Writing – review & editing.

**Declaration of competing interest**
The authors declare that they have no known competing financial interests or personal relationships that could have appeared to influence the work reported in this paper.

**Acknowledgments**
This work is supported, in part, by the Polymers program at the USA National Science Foundation (DMR-1905870, 2210184). We are grateful to Dr. Gregory Brust at Lion Elastomers for providing us the EPDM sheets.

**References**

1. Creton, C.; Ciccotti, M. Fracture and adhesion of soft materials: a review. *Rep. Prog. Phys.* **2016,** 79, (4).
2. Long, R.; Hui, C.-Y.; Gong, J. P.; Bouchbinder, E. The fracture of highly deformable soft materials: A tale of two length scales. *Annu. Rev. Condens. Matter Phys.* **2021,** 12, 71-94.
3. Griffith, A. A. VI. The phenomena of rupture and flow in solids. *Philosophical transactions of the royal society of london. Series A, containing papers of a mathematical or physical character* **1921,** 221, (582-593), 163-198.
4. Anderson, T. L., *Fracture mechanics: fundamentals and applications*. CRC press: 2017.
5. Kinloch, A. J., *Fracture behaviour of polymers*. Springer Science & Business Media: 2013.
6. Berry, J. Fracture processes in polymeric materials. I. The surface energy of poly (methyl methacrylate). *Journal of Polymer Science* **1961,** 50, (153), 107-115.
7. Berry, J. Fracture processes in polymeric materials. II. The tensile strength of polystyrene. *Journal of Polymer Science* **1961,** 50, (154), 313-321.
8. Rivlin, R.; Thomas, A. G. Rupture of rubber. I. Characteristic energy for tearing. *Journal of polymer science* **1953,** 10, (3), 291-318.
9. Inglis, C. E. Stresses in a plate due to the presence of cracks and sharp corners. *Trans Inst Naval Archit* **1913,** 55, 219-241.
10. Westergaard, H. M. Bearing pressures and cracks. *Trans AIME, J. Appl. Mech.* **1939,** 6, 49-53.
11. Irwin, G. R. Analysis of stresses and strains near the end of a crack transversing a plate. *Trans. ASME, Ser. E, J. Appl. Mech.* **1957,** 24, 361-364.
12. Sun, C.-T.; Jin, Z., *Fracture mechanics*. Academic Press: 2011.
13. Williams, M. L. On the stress distribution at the base of a stationary crack. *J. Appl. Mech.* **1957,** 24, 109-114.
14. Irwin, G. R. *Onset of fast crack propagation in high strength steel and aluminum alloys*; Sagamore Research Conference Proceedings: 1956; pp 289-305.
15. Irwin, G. Relation of stresses near a crack to the crack extension force. *9th Cong. App. Mech., Brussels* **1957**.
16. Smith, T.; Gupta, C.; Fan, Z.; Brust, G. J.; Vogelsong, R.; Carr, C.; Wang, S.-Q. Toughness arising from inherent strength of polymers. *Extreme Mechanics Letters* **2022**, 101819.
17. Fraser, R. A. W.; Ward, I. M. Temperature dependence of craze shape and fracture in polycarbonate. *Polymer* **1978,** 19, (2), 220-224.
18. Lake, G. J. High-Speed Fracture of Elastomers: Part I. *Rubber Chemistry and Technology* **2000,** 73, (5), 801–817.
19. Thomas, A. K. A. G. Tear Behavior of Rubbers Over a Wide Range of Rates. *Rubber Chemistry and Technology* **1981,** 54, (1), 15–23.
20. Gent, A.; Lai, S.; Nah, C.; Wang, C. Viscoelastic effects in cutting and tearing rubber. *Rubber Chem. Technol.* **1994,** 67, (4), 610-618.
21. Gent, A. Adhesion and strength of viscoelastic solids. Is there a relationship between adhesion and bulk properties? *Langmuir* **1996,** 12, (19), 4492-4496.
22. Persson, B.; Albohr, O.; Heinrich, G.; Ueba, H. Crack propagation in rubber-like materials. *J. Phys.: Condens. Matter* **2005,** 17, (44), R1071.
23. Creton, C. 50th anniversary perspective: Networks and gels: Soft but dynamic and tough. *Macromolecules* **2017,** 50, (21), 8297-8316.
24. Mullins, L. Rupture of rubber. IX. Role of hysteresis in the tearing of rubber. *Transactions of the Institution of the Rubber Industry* **1958,** 35, (5), 213-222.
25. Greensmith, H. Rupture of rubber. X. The change in stored energy on making a small cut in a test piece held in simple extension. *J. Appl. Polym. Sci.* **1963,** 7, (3), 993-1002.

26. Creton, C.; Ciccotti, M. Fracture and adhesion of soft materials: a review. *Rep. Prog. Phys.* **2016,** 79, (4), 046601.
27. Lake, G.; Thomas, A. The strength of highly elastic materials. *Proc. R. Soc. Lond. A* **1967,** 300, (1460), 108-119.
28. Thomas, A. Rupture of rubber. II. The strain concentration at an incision. *J. Polym. Sci.* **1955,** 18, 177-188.
29. Ducrot, E.; Chen, Y. L.; Bulters, M.; Sijbesma, R. P.; Creton, C. Toughening Elastomers with Sacrificial Bonds and Watching Them Break. *Science* **2014,** 344, (6180), 186-189.
30. Glinka, G. Calculation of inelastic notch-tip strain-stress histories under cyclic loading. *Engineering Fracture Mechanics* **1985,** 22, (5), 839-854.
31. Glinka, G.; Newport, A. Universal features of elastic notch-tip stress fields. *Int. J. Fatigue* **1987,** 9, (3), 143-150.
32. Lazzarin, P.; Tovo, R. A unified approach to the evaluation of linear elastic stress fields in the neighborhood of cracks and notches. *Int J Fract* **1996,** 78, (1), 3-19.
33. Filippi, S.; Lazzarin, P.; Tovo, R. Developments of some explicit formulas useful to describe elastic stress fields ahead of notches in plates. *International Journal of Solids and Structures* **2002,** 39, (17), 4543-4565.